\documentclass[prd,12pt,superscriptaddress,preprintnumbers]{revtex4}
\usepackage{epsfig}
\usepackage{graphicx}
\usepackage{amsmath}
\usepackage{amsfonts}
\usepackage{amssymb}
\usepackage{amsbsy}
\usepackage{pstricks}
\newcommand{\AHEP}{Instituto de F\'{\i}sica Corpuscular --
  C.S.I.C./Universitat de Val{\`e}ncia \\
  Edificio Institutos de Paterna, Apt 22085,
  E--46071 Val{\`e}ncia, Spain\\}
\begin{document} 

\title{Probing neutrino mass with displaced vertices at the Tevatron }

%{in the Broken R--parity Minimal Supersymmetric Standard Model}

\author{F.\ de Campos}
\email{fernando@ift.unesp.br}
\affiliation{Departamento de F\'{\i}sica e Qu\'{\i}mica,
             Universidade Estadual Paulista, Guaratinguet\'a -- SP,  Brazil }

\author{O.\ J.\ P.\ \'Eboli}
\email{eboli@fma.if.usp.br}
\affiliation{Instituto de F\'{\i}sica, 
             Universidade de S\~ao Paulo, S\~ao Paulo -- SP, Brazil.}

\author{M.\ B.\ Magro}
\email{magro@fma.if.usp.br}
\affiliation{Instituto de F\'{\i}sica, 
             Universidade de S\~ao Paulo, S\~ao Paulo -- SP, Brazil.}
           
\author{W.\ Porod} 
\email{porod@ific.uv.es} 
\affiliation{\AHEP}
\affiliation{Inst.~f\"ur Theoretische Physik, Uni.~Z\"urich }

\author{D.\ Restrepo}
\email{restrepo@uv.es}
\affiliation{\AHEP}
\affiliation{Instituto de F\'{\i}sica, Universidad de Antioquia - Colombia}

\author{J.\ W.\ F.\ Valle} 
\email{valle@ific.uv.es}
\affiliation{\AHEP}

%%%%%%%%%%%%%%%%%%%%%%%%%%%%%%%%%%%%%%%%%%%%%%%%%%%%%%%
     
\begin{abstract}

\vskip 36pt

Supersymmetric extensions of the standard model exhibiting bilinear R--parity
violation can generate naturally the observed neutrino mass spectrum as well
mixings. One interesting feature of these scenarios is that the lightest
supersymmetric particle (LSP) is unstable, with several of its decay
properties predicted in terms of neutrino mixing angles.  A smoking gun of
this model in colliders is the presence of displaced vertices due to LSP
decays in large parts of the parameter space.  In this work we focus on the
simplest model of this type that comes from minimal supergravity with
universal R--parity conserving soft breaking of supersymmetry (RmSUGRA). We
evaluate the potentiality of the Fermilab Tevatron to probe the RmSUGRA
parameters through the analysis of events possessing two displaced vertices
stemming from LSP decays.  We show that requiring two displaced vertices in
the events leads to a reach in $m_{1/2}$ twice the one in the usual
multilepton signals in a large fraction of the parameter space.

\medskip

%{\bf draft6.tex -- 17/01/2005}

\end{abstract}

\preprint{IFIC/04-72, ZU-TH 04/03 }

\maketitle
%\newpage

%%%%%%%%%%%%%%%%%%%%%%%%%%%%%%%%%%%%%%%%%%%%%%%%%%%%%%%%%%%%%%%%%%%%%%

\section{Introduction}

Despite the lack of direct experimental evidence, supersymmetry is the
most popular candidate for physics beyond the standard model (SM).
Apart from its known motivations related in particular to the
hierarchy problem, supersymmetry may provide also the understanding of
the origin of neutrino mass~\cite{Hirsch:2004he}. Supersymmetric
(SUSY) models have been used as benchmarks to evaluate the
potentiality for new discoveries at the current high energy colliders,
such as the Tevatron, and the future colliders, like the CERN Large
Hadron Collider or the International Linear Collider.  Collider
signals for supersymmetry have been studied both in the context of
R--parity conservation~\cite{susywg,tevatron,lhc,epem}, 
as well as in scenarios with
R--parity
violation~\cite{Allanach:1999bf,displacedvertices,Magro:2003zb,Barger:1994nm,Jung:2004rd}.

A variety of recent neutrino physics experiments~\cite{Maltoni:2004ei}
has established the existence of neutrino oscillations and masses,
indicating clearly the need for physics beyond the SM.  It is tempting
to imagine that neutrino physics and supersymmetry are tied
together~\cite{rpold}: SUSY models exhibiting R--parity violation can
lead to a pattern of neutrino masses and mixings~\cite{Hirsch:2000ef}
in agreement with the current solar and reactor neutrino data as well
as atmospheric and accelerator neutrino data.  The simplest
realization of this idea assumes bilinear violation of
R--parity~\cite{bilinear,Diaz:1997xc}.  Such restricted R--parity
violation can arise either as an effective description of a
spontaneous R--parity violation scenario~\cite{Masiero:1990uj}, or as
a result of suitable symmetries of the underlying
theory~\cite{Bento:1987mu}.  It also corresponds to the simplest
broken R--parity version of the Minimal Supersymmetric Standard Model
(MSSM), to which we refer, in what follows, as RMSSM.

Despite the presence of additional interactions in R--parity violating
supersymmetry, the SUSY particle production cross sections at colliders are
essentially the same as in the MSSM, since the smallness of neutrino masses
implies small R--parity violation parameters.  Similarly, the SUSY particle
cascade decays are controlled by R--parity conserving interactions until the
LSP is produced.  In contrast to R--parity conserving supersymmetry, the LSP
is unstable in these models as it is not protected by any symmetry. In fact,
R--parity violating interactions determine its branching ratios and can lead
to large decay lengths and displaced vertices if these interactions are
sufficiently small.  Parenthetically, let us mention that a light gravitino,
as it occurs for example gauge mediated SUSY breaking, is a possible candidate
for dark matter ~\cite{Takayama:2000uz}. In case that the gravitino is heavy,
dark matter must come from some other source, like the QCD--sector axion.  In
the light gravitino case displaced vertices discussed here should come from
the R--parity violating NLSP decays which still have a large
probability~\cite{Hirsch:2003fe}.  Turning to the case of interest, namely the
RMSSM, it has been shown that in this case the LSP decay properties can be
directly tested at collider
experiments~\cite{Bartl:2000yh,Porod:2000hv,Hirsch:2003fe,Hirsch:2002ys,displacedvertices,Magro:2003zb,otherbrpv}.

In this work, we focus on the possibility of the $\tilde{\chi}^0_1$ decay
length being large enough to generate displaced vertices at the Tevatron,
which could be observed by the tracking systems of the experimental
collaborations.  We work within the framework of the simplest model of this
kind, namely the minimal supergravity (mSUGRA) version of the RMSSM with
universal soft breaking terms at unification~\cite{Hirsch:2000ef}. We call it
RmSUGRA, for short.  However, here we are not enforcing universality
conditions for R--parity violating soft terms.  In such models one can show
that over a large range of the accessible parameter space the masses of the
sleptons, the lighter chargino ($\tilde{\chi}^\pm_1$), and the lighter
neutralinos ($\tilde{\chi}^0_1$ and $\tilde{\chi}^0_2$) are considerably
smaller than the gluino and squark masses; see {\em e.g.}~\cite{tata1} and
references therein.  As a result, at Tevatron energies, the largest reach in
supersymmetry searches comes from the production of charginos and neutralinos,
and one might expect a copious number of displaced vertex events at the
Tevatron depending on the parameters of the model~\cite{displacedvertices}.

In this paper we show that the presence of bilinear R--parity
violating (BRpV) interactions enhances the discovery reach for
supersymmetry over most of the parameter space with respect to the one
obtained using the trilepton/multilepton
signal~\cite{chung,Magro:2003zb}.  Previous analysis of the potential
of the Tevatron for probing neutrino mass models based on R--parity
violation either analyzed the changes in the multilepton signal due to
the LSP decay~\cite{Magro:2003zb} or studied the same sign dilepton
production presenting a displaced vertex due to the decay of a
neutralino into $\mu W$ or $\tau W$ pairs~\cite{displacedvertices}.
Moreover, these analyses were performed imposing only the constraints
from atmospheric neutrino data.  Here, we analyzed events exhibiting
two reconstructed displaced vertices and we scan the full accessible
RmSUGRA parameter space at Tevatron, taking into account all
constraints coming from  neutrino masses and mixing angles in the BRpV
model. We show that the predicted number of events at Tevatron depends
mainly on $m_{1/2}$, or equivalently the neutralino mass, with mild
dependences on the other mSUGRA parameters or specific neutralino
decay channels.

This paper is organized as follows: Section \ref{model} summarizes the
most relevant features of the RmSUGRA model. The displaced vertex
analysis is presented in section \ref{cuts}, while we summarize our
conclusions in section \ref{concl}.

%%%%%%%%%%%%%%%%%%%%%%%%%%%%%%%%%%%%%%%%%%%%%%%%%%%%%%%%%%%%%%%%%%%%%%
\section{Main features of the RmSUGRA model}
\label{model}

Here we consider a simple variant of the MSSM that includes the
following bilinear terms in the
superpotential~\cite{bilinear,Diaz:1997xc,bilinear1}
\begin{equation}
W_{\text{BRpV}} = W_{\text{MSSM}}  + \varepsilon_{ab}
\epsilon_i \widehat L_i^a\widehat H_u^b \; ,
\end{equation}
The relevant bilinear terms in the soft supersymmetry breaking sector
are
\begin{equation}
V_{\text{soft}} = m_{H_u}^2H_u^{a*}H_u^a+m_{H_d}^2H_d^{a*}H_d^a+
M_{L_i}^2\widetilde L_i^{a*}\widetilde L_i^a -\varepsilon_{ab}\left(
B\mu H_d^aH_u^b+B_i\epsilon_i\widetilde L_i^aH_u^b\right) \; ,
\end{equation}
The simultaneous presence of the bilinear term in the superpotential
and its analogue in the soft supersymmetry breaking sector violates
R--parity and lepton number explicitly, inducing vacuum expectation
values (vev) $v_i$, $i=1,2,3$ for the sneutrinos.  In order to
reproduce the observed values of neutrino masses and
mixings~\cite{Maltoni:2004ei} we must have $|\epsilon_i| \ll |\mu|$,
where $\mu$ denotes the SUSY bilinear mass
parameter~\cite{Hirsch:2000ef}.

In a minimal supergravity model with universal soft breaking terms at
unification the relevant parameters are
\begin{equation}
m_0\,,\, m_{1/2}\,,\, \tan\beta\,,\, {\mathrm{sign}}(\mu)\,,\, 
A_0 \,,\, 
\epsilon_i \: {\mathrm{, and}}\,\, \Lambda_i\,,
\end{equation}
where $m_{1/2}$ and $m_0$ are the common gaugino mass and scalar soft
SUSY breaking masses at the unification scale, $A_0$ is the common
trilinear term, and $\tan\beta$ is the ratio between the Higgs field
vev's.  Although many parameterizations are possible, we take as the
free parameters the $\epsilon_i$ of the superpotential and the
so-called alignment parameters, $\Lambda_i=\epsilon_iv_d+\mu v_i$. Moreover, we
are not enforcing universality conditions for R--parity violating soft
terms.

Successful explanation of neutrino data requires small R--parity
violating couplings. Therefore, the induced shift in the SUSY masses
by the BRpV interactions is smaller than the foreseeable experimental
accuracy, and thus, we use the following strategy: for a given set of
the R--parity conserving parameters we calculate the mass spectrum
setting the R--parity violating parameters to zero when evolving the
parameters from the unification scale to the electroweak scale.  This
procedure was carried out using the package SPheno; details of this
calculation can be found in~\cite{Porod:2003um}.  At the electroweak
scale we choose the R--parity violating parameters such that the
neutrino data are successfully explained within the $3\sigma$ range as
given,{\em e.g.}~in Ref.~\cite{Maltoni:2004ei}:
\begin{eqnarray}
0.23 \lesssim & \sin^2\theta_{\mathrm{sol}} & \lesssim 0.38 \; , 
\\
7.1~\times10^{-5} \,{\mathrm{eV}}^2 \lesssim & \Delta m^2_{\mathrm{sol}} & \lesssim 
8.9~\times10^{-5} \,{\mathrm{eV}}^2  \; ,
\\
0.34 \lesssim & \sin^2\theta_{\mathrm{atm}} & \lesssim 0.68 \; , 
\\
1.4\times10^{-3} \,{\mathrm{eV}}^2 \lesssim & \Delta m^2_{\mathrm{atm}} & \lesssim 
3.3\times10^{-3}
\,{\mathrm{eV}}^2  \; .
\label{neut:atm}
\end{eqnarray}
In order to obtain the mixing matrices between SM and SUSY particles
we diagonalize the corresponding mass matrices containing SM and SUSY
fermions and scalars, for instance, the full $7\times7$ mass matrix
for neutrinos and neutralinos.  Afterwards all SUSY R--parity
violating as well as R--parity conserving decays are calculated using
an adapted version of SPheno.  The output of this calculation is
tabulated in the SLHA format ~\cite{Skands:2003cj} and inputed into
PYTHIA~\cite{pythia}, which is used to generate the events.

\begin{figure}[thbp] 
  \includegraphics[scale=0.5]{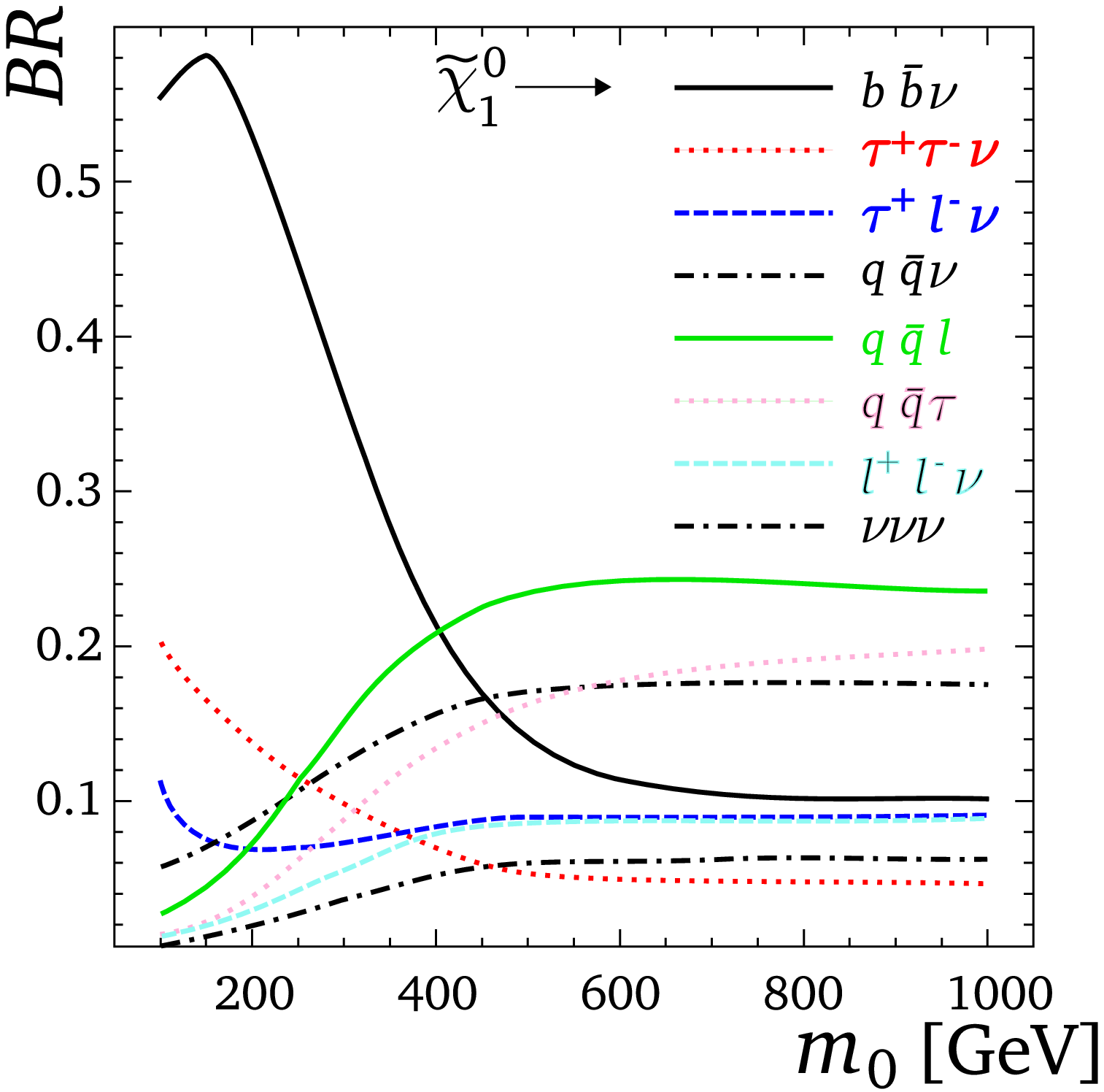}   
  \caption{$\tilde\chi_1^0$ branching ratios versus $m_0$ for $A_0=-100$ GeV,
    $m_{1/2} =350$ GeV, $\tan\beta=10$ and $\mu > 0$.  We took the R--parity
    violating parameters for each mSUGRA point such that they are compatible
    with neutrino data and the neutralino decay length is maximized.  }
\label{fig:br}
\end{figure}

We studied the scenarios where the LSP is the lightest neutralino
which decays into SM particles due to the mixings induced by the BRpV
interactions. The lightest neutralino has leptonic decays
$\tilde{\chi}^0_1 \to \nu \ell^+ \ell^{\prime-}$, the invisible mode
$\tilde{\chi}^0_1 \to \nu \nu \nu$ and semi-leptonic decays
$\tilde{\chi}^0_1 \to \nu q \bar{q} $, $\ell q
\bar{q}$~\cite{Porod:2000hv}.  As an illustration, we depict in Fig.\
\ref{fig:br} the lightest neutralino branching ratios as a function of
$m_0$ for $m_{1/2} =350$ GeV, $A_0 = -100$ GeV, $\mu > 0$, and
$\tan\beta = 10$.  In this plot we took the R--parity violating
parameters for each mSUGRA point that lead to the largest decay length
compatible with neutrino data at the $3\sigma$ level.  As we can see
from this figure, the decay $\tilde{\chi}^0_1 \to \nu b \bar{b}$
dominates at small $m_0$. This decay channel arises from an effective
coupling $\lambda^\prime_{i33}\sim (\epsilon_i/\mu)h_b$, induced by
the Higgs-slepton mixing, where $h_b$ is the bottom Yukawa coupling.
Therefore, this decay is enhanced at small $m_0$ due to the lightness
of the scalars mediating it and at large $\tan\beta$ due to enhanced
Yukawa couplings. Moreover, at large $m_0$ all $\tilde{\chi}^0_1$
decay channels give a sizeable contribution with the main decay modes
being the semi-leptonic ones exhibiting light quarks.

\begin{figure}[thbp] 
  \includegraphics[scale=0.5]{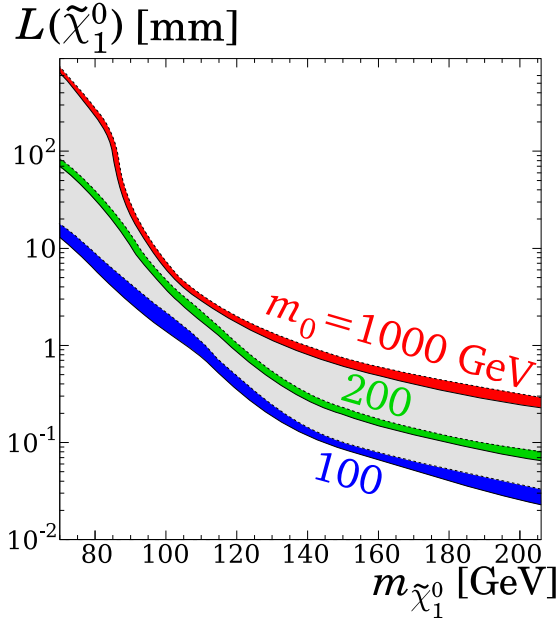}   
  \caption{ $\tilde\chi_1^0$ decay length versus $m_{1/2}$ for
    $A_0=-100$ GeV, $\tan\beta=10$ and $\mu > 0$. The width of the
    $m_0$ bands is due to the variation of the BRpV parameters in such
    a way that the neutrino masses and mixing angles are within
    $3\sigma$ of their best fit values.}
\label{fig:dpath}
\end{figure}

The expected $\tilde{\chi}^0_1$ lifetime (decay length) depends both
on the magnitude of R--parity breaking parameters and the chosen
values of the mSUGRA parameters.  The smallness of the R--parity
violating parameters implies that the lightest neutralino has a small
decay width~\cite{Porod:2000hv}, thus having a lifetime large
enough to give rise to displaced vertices.  In Fig.\ \ref{fig:dpath}
we depict the lightest neutralino decay length as a function of
$m_{1/2}$ for $m_0=$ 100, 200, and 1000 GeV and $A_0 = -100$, $\mu >
0$, $\tan\beta = 10$. In this figure, we varied the R--parity
violating parameters so that $|\vec \epsilon/\mu|$ and $|\vec
\Lambda|^2$ lie in the $3\sigma$ bands required to account for the
neutrino oscillation data~\cite{Maltoni:2004ei}.  Note from Fig.\
\ref{fig:dpath} that the LSP decay length strongly depends on $m_0$,
as it controls the masses of the scalar particles exchanged in the LSP
decay.  In contrast, the dependence on $A_0$ is rather mild. One can
see from this figure that the LSP decay length is in the range of
tenths of mm up to $\simeq 10$ cm.  Therefore, the lightest neutralino
should decay inside the detectors of Tevatron yielding an observable
displaced vertex.

%%%%%%%%%%%%%%%%%%%%%%%%%%%%%%%%%%%%%%%%%%%%%%%%%%%%%%%%%%%%%%%%%%%%%%
\section{Signal, Backgrounds, and Selection Cuts}
\label{cuts}

In the context of our RMSSM/RmSUGRA, supersymmetric particle
production and subsequent cascade decays proceed as in the R--parity
conserving case (MSSM/mSUGRA) up to the final state containing two
lightest neutralinos, as mentioned above.  Since neutrino data require
the R--parity violating couplings to be small, the decays of these two
neutralinos leads, in general, to displaced
vertices~\cite{Porod:2000hv}.  Assuming that gluinos and squarks are
too heavy to be produced at the Tevatron, the most important SUSY
production processes at the Tevatron are
\[
p \bar{p} \to \tilde{\ell}^+ \tilde{\ell}^- \;\;\;,\;\;\; 
\tilde{\nu} \tilde{\ell} \;\;\; , \;\;\;
\tilde{\chi}^0_i \tilde{\chi}^0_j~ (i(j)=1,2) \;\;\;,\;\;\; 
\tilde{\chi}^+_1 \tilde{\chi}^-_1 \;\;\;\hbox{,  and} \;\;\;
\tilde{\chi}^0_i \tilde{\chi}^\pm_1~ (i=1,2) \;.
\]

The main decay modes of the lightest neutralino in our model are
%\begin{enumerate}
\begin{itemize}
  
\item $\tilde{\chi}^0_1 \to \nu \ell^+ \ell^-$ with $\ell =e$, $\mu$
  denoted by $\ell \ell$;

  \item $\tilde{\chi}^0_1 \to \nu q \bar{q}$ denoted $jj$;

  \item $\tilde{\chi}^0_1 \to \tau q^\prime \bar{q}$, called $\tau jj$;
    
  \item $\tilde{\chi}^0_1 \to \nu b \bar{b}$, that we denote by $bb$;

  \item $\tilde{\chi}^0_1 \to \nu \tau^+ \tau^-$, called $\tau \tau$;

  \item  $\tilde{\chi}^0_1 \to \tau \nu  \ell$, called $\tau \ell$.

\end{itemize}
%\end{enumerate}

In order to mimic the triggers used by the Tevatron collaborations, we 
accept events passing at least one of the following requirements:
\begin{enumerate}

  \item the event has two muons with transverse momenta satisfying 
    $p_T > 4$ GeV;
  
  \item the event possesses a lepton (electron or muon) with $p_T> 4$ GeV and
    a displaced track with an impact parameter in excess of 0.120 mm and $p_T
    > 2 $ GeV;

   \item the event exhibits two displaced tracks of opposite charge with
      $p_T> 3$ GeV and impact parameter larger than 0.120 mm.

\end{enumerate}

In the $\ell \ell$, $jj$, $\tau jj$, and $bb$ decays, the tracks point
to the LSP decay vertex, up to measurement errors. In our analysis we
smeared the energies, but not directions, of all final state particles
with a Gaussian error given by $\Delta E/E = 0.7/\sqrt{E}$ ($E$ in
GeV) for hadrons and $\Delta E /E = 0.15/\sqrt{E}$ for charged
leptons. Therefore, the LSP decay vertex is reconstructed correctly
for these modes within these approximations.  On the other hand, the
tracks originating from the $\tau \tau$ and $\tau \ell$ decay modes do
not converge to the LSP decay vertex. For these events we looked for
tracks that do not point to the interaction region and whose
reconstructed trajectories come closer than $100~\mu$m ($ 1000~\mu$m)
if the closest point of the two trajectories are inside the silicon
detector (central tracker). In this case, we defined the position of
the displaced vertex as the average of the closest points of the two
trajectories.

In our analysis, we also required that the two reconstructed vertices
are away from the interaction point, more specifically, we demanded
that both LSP decay vertices lie outside an ellipsoid
\[
      \left ( \frac{x}{\delta_{xy}} \right )^2
   +  \left ( \frac{y}{\delta_{xy}} \right )^2
   +  \left ( \frac{z}{\delta_{z}} \right )^2   = 1 \; ,
\] 
where the $z$-axis is along the beam direction. We assumed that $\delta_{xy} =
150~\mu$m and $\delta_z = 300~\mu$m.

To guarantee a high efficiency in the reconstruction of the displaced
vertices without a full detector simulation, we restricted our
analysis to reconstructed vertices with pseudo-rapidities $| \eta| <
1.5$.  Moreover, both neutralinos should decay well inside the inner
detectors to guarantee enough hits in the track system. Therefore, we
require that the neutralinos decay inside a cylinder with radius
$r=350$ mm and length $l=1250$ mm.

\begin{figure}[htbp]
  \includegraphics[scale=0.5]{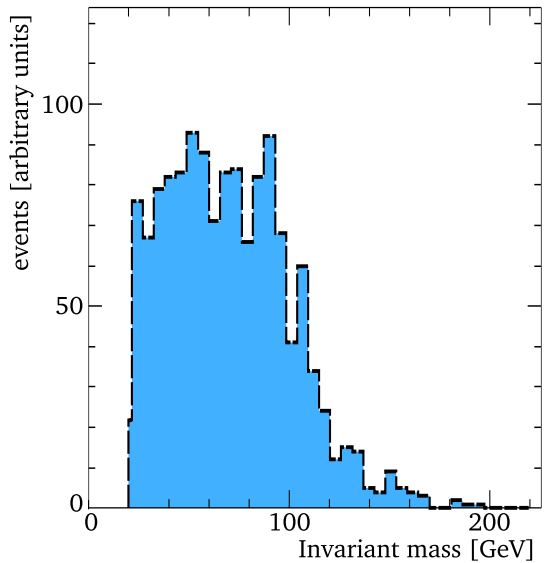}
  \caption{Reconstructed invariant mass spectrum of displaced vertices after
    applying the cuts in Sec.~\ref{cuts}. Here we chose $m_{1/2}=350$ GeV,
    $m_0=400$ GeV, $A_0=-100$ GeV, $\mu > 0$, $\tan\beta=10$, and the
    R--parity violating parameters for each mSUGRA point such that they are
    compatible with neutrino data and the neutralino decay length is
    maximized. }
  \label{fig:minva}
\end{figure}

The SM backgrounds coming, for instance, from displaced vertices
associated to $b$'s or $\tau$'s can be eliminated by requiring that
the set of tracks defining a displaced vertex should have an invariant
mass larger than 20~GeV.  This way the displaced vertex signal passing
all the above cuts is essentially background free.
Fig.~\ref{fig:minva} displays the reconstructed invariant mass of
displaced vertices after cuts for $m_0 = 400$ GeV, $m_{1/2} =350$ GeV,
$A_0 = -100$ GeV, $\mu > 0$, $\tan\beta = 10$ and
$|\boldsymbol{\epsilon} | /\mu$ taking the largest value compatible
with neutrino data.  As we can see, the neutralino decays lead to a
substantial invariant mass associated to the displaced vertices.  If
necessary, the invariant mass can be further tightened without a
substantial loss of signal.

\begin{figure}[htbp]
  \includegraphics[scale=0.5]{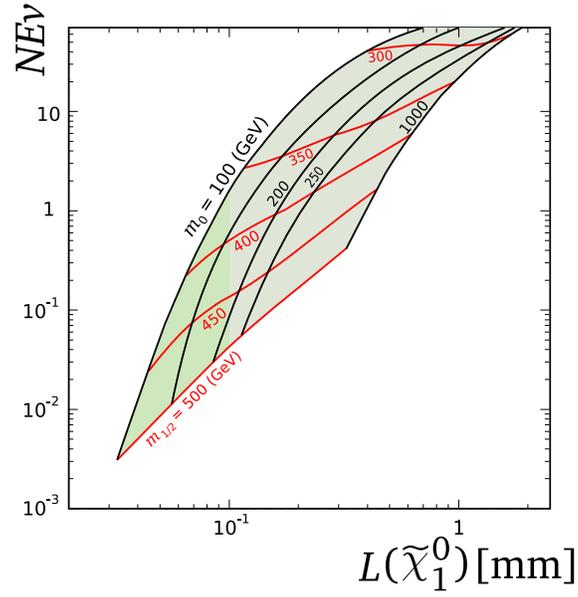}
  \caption{Expected number of events for an integrated luminosity of 8
    $\text{fb}^{-1}$ as a function of the neutralino decay length for
    $A_0=-100$ GeV, $\mu > 0$, $\tan\beta=10$, and the R--parity violating
    parameters for each mSUGRA point such that they are compatible with
    neutrino data and the neutralino decay length is maximized. }
  \label{fig:nevdl}
\end{figure}

We display in Fig.~\ref{fig:nevdl} the expected number of events ($NEv$) after
the above cuts as function of the neutralino decay length for various values
of $m_0$ and $m_{1/2}$ and an integrated luminosity of 8 fb$^{-1}$.  In this
plot we chose the R--parity violating parameters such that the largest decay
length is obtained compatible with neutrino data at the $3\sigma$ level. The
same plot for smaller decay lengths is basically the same but with a shift of
large $m_0$ values to the left side.  For example, for the point with
$(m_{1/2},m_0)=(350,1000)\,$GeV we have $NEv=21$, $L(\tilde\chi_1^0) = 0.9
\,$mm and $|\boldsymbol{\epsilon} / \mu|=5.2 \times10^{-4}$, while for smaller
value of $|\boldsymbol{\epsilon} / \mu|=4\times10^{-4}$ for the same SUGRA
point, we find $NEv=$ 18.4 and $L(\tilde\chi_1^0)=0.74\,$mm.  In general, the
variation of the number of events at any SUGRA point is of order 10\% for the
full range for the R--parity parameters compatible with neutrino data.  The
reason for this behaviour is that the decay length is mainly determined by the
values of the neutrino masses which are already known rather precisely.

\begin{figure}[htbp]
  \includegraphics[scale=0.5]{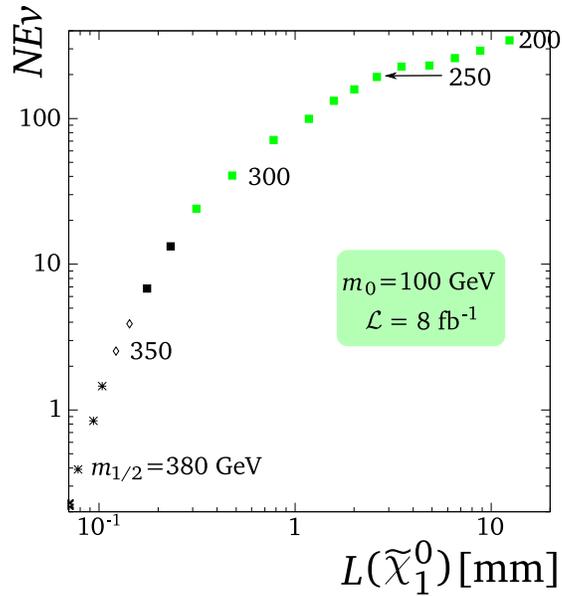}
  \caption{Same as Fig.~\protect{\ref{fig:nevdl}} but fixing $m_0 =
    100\,$GeV.  }
  \label{fig:nevdl0100}
\end{figure}

In order to illustrate the behavior of the signal for a fixed value of
$m_0$ we present, in Fig.~\ref{fig:nevdl0100}, the expected number of
events for the same parameters as in Fig.~\ref{fig:nevdl} but for
fixed $m_0 = 100\,$ GeV.  The numbers given beside some points
indicate the corresponding value of $m_{1/2}$.  At small decay lengths
($\lesssim 0.3\,$mm) the production cross section of neutralinos is
rather low, as well as the efficiency for extracting the signal. As
the decay length increases, $m_{1/2}$ ($m_{\tilde{\chi}^0_1}$)
decreases and the production cross section increases, however, the
detection efficiency reaches a maximum around $\simeq 1\,$mm and then
decreases again.  These two opposite behaviors lead to an
approximately constant signal rate for decay lengths $\gtrsim 10$ mm.

\begin{figure}[htbp]
 \includegraphics[scale=0.4]{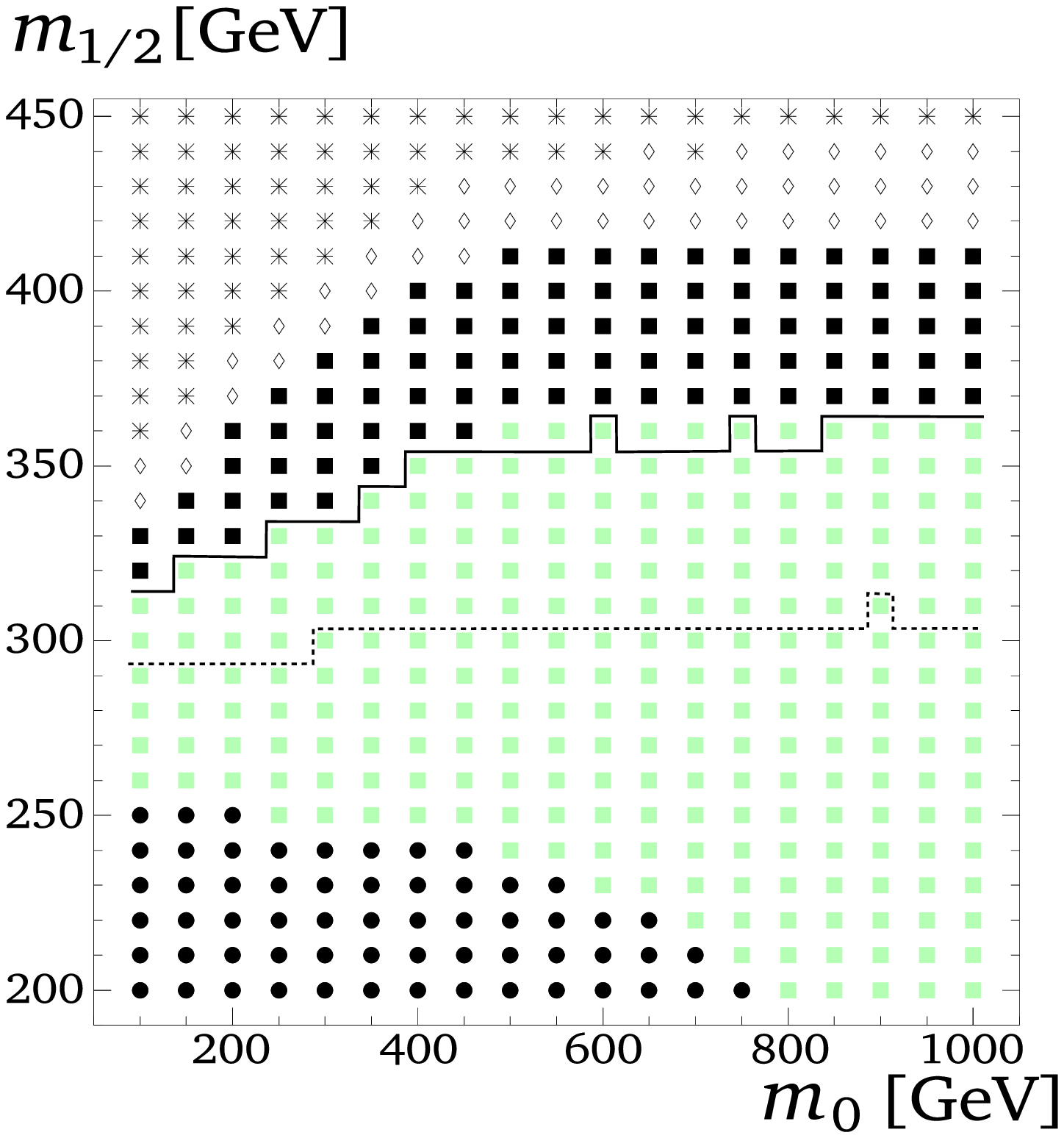}
 \includegraphics[scale=0.4]{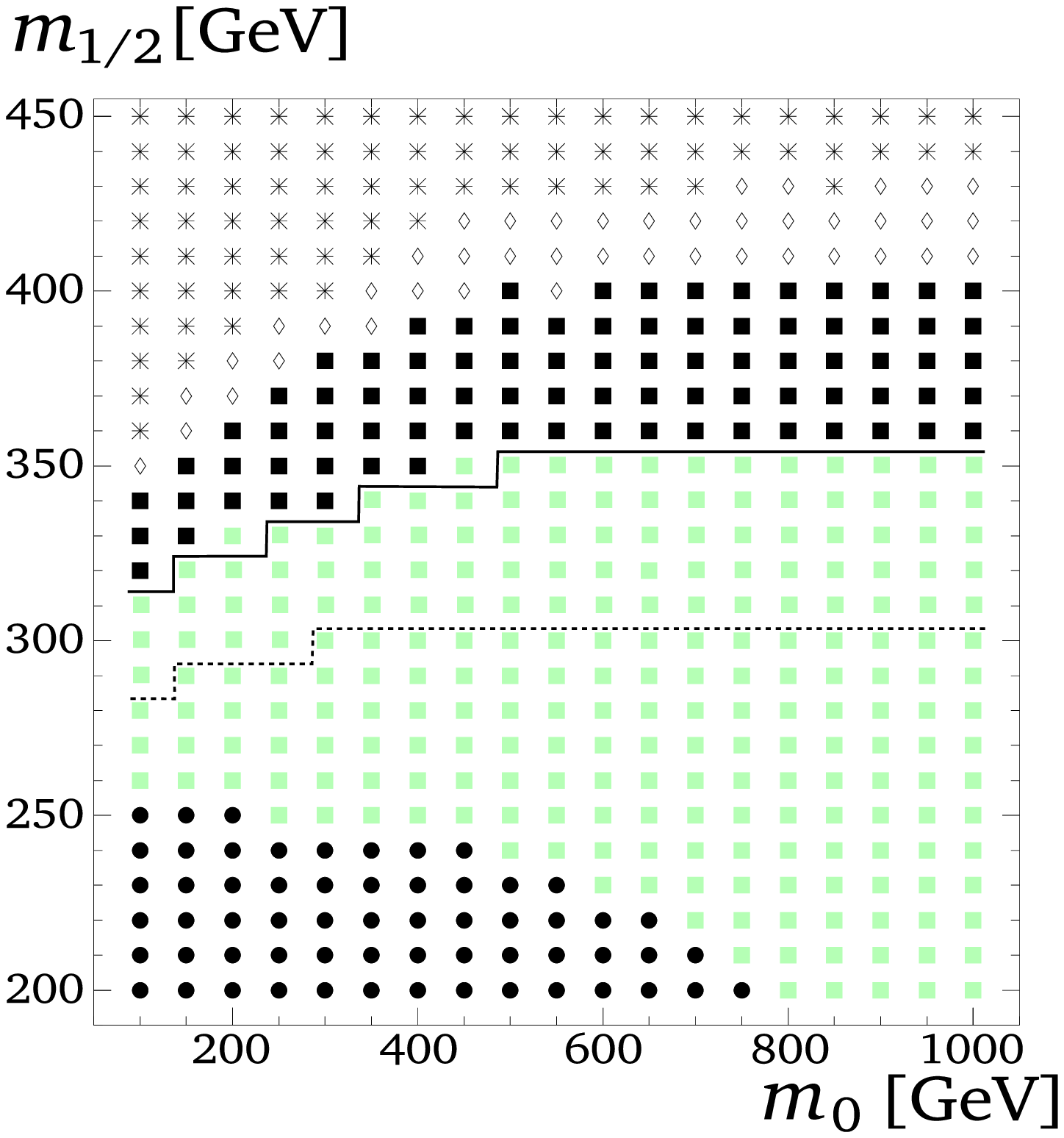}\\
 \includegraphics[scale=0.4]{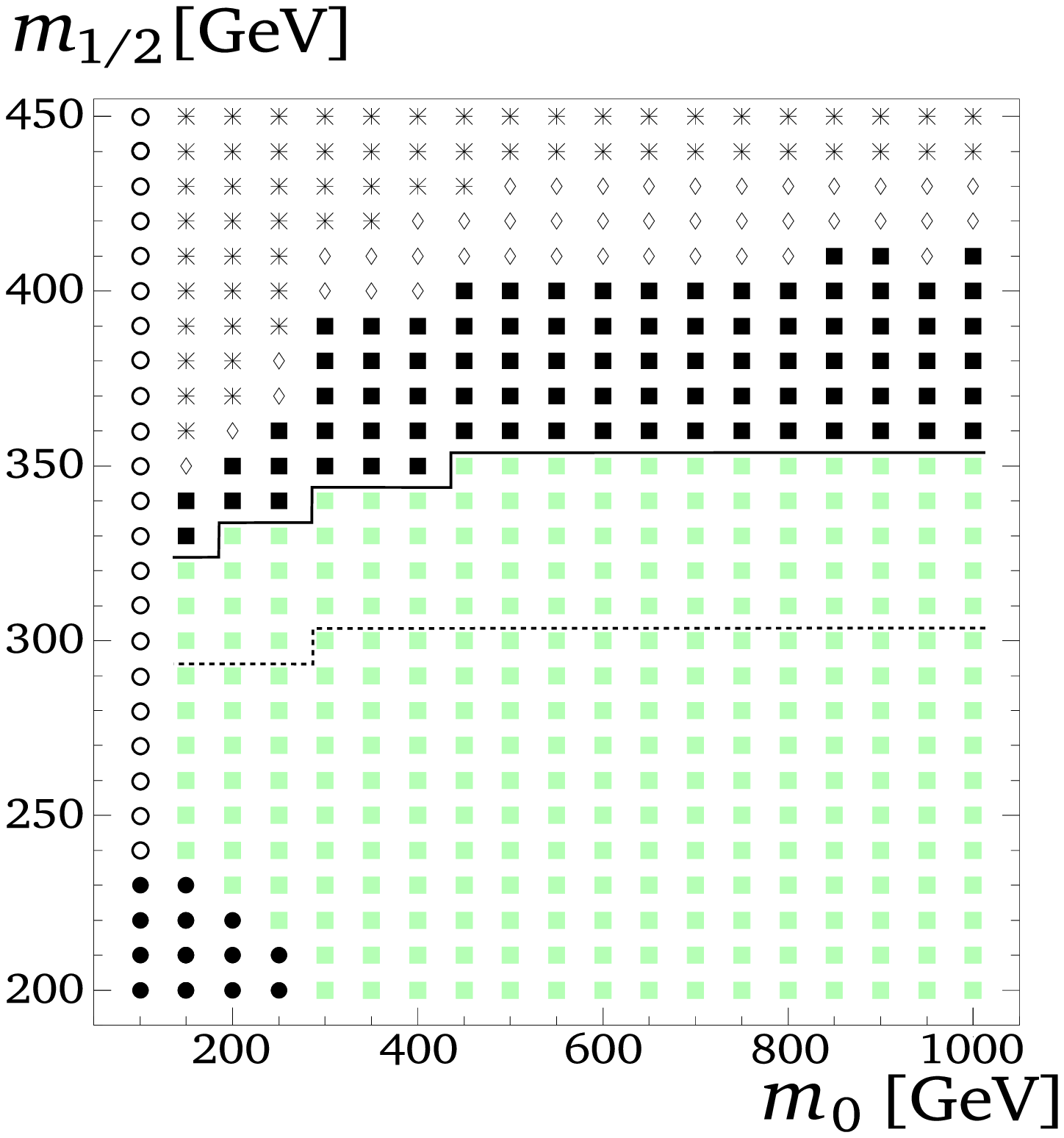}
 \includegraphics[scale=0.4]{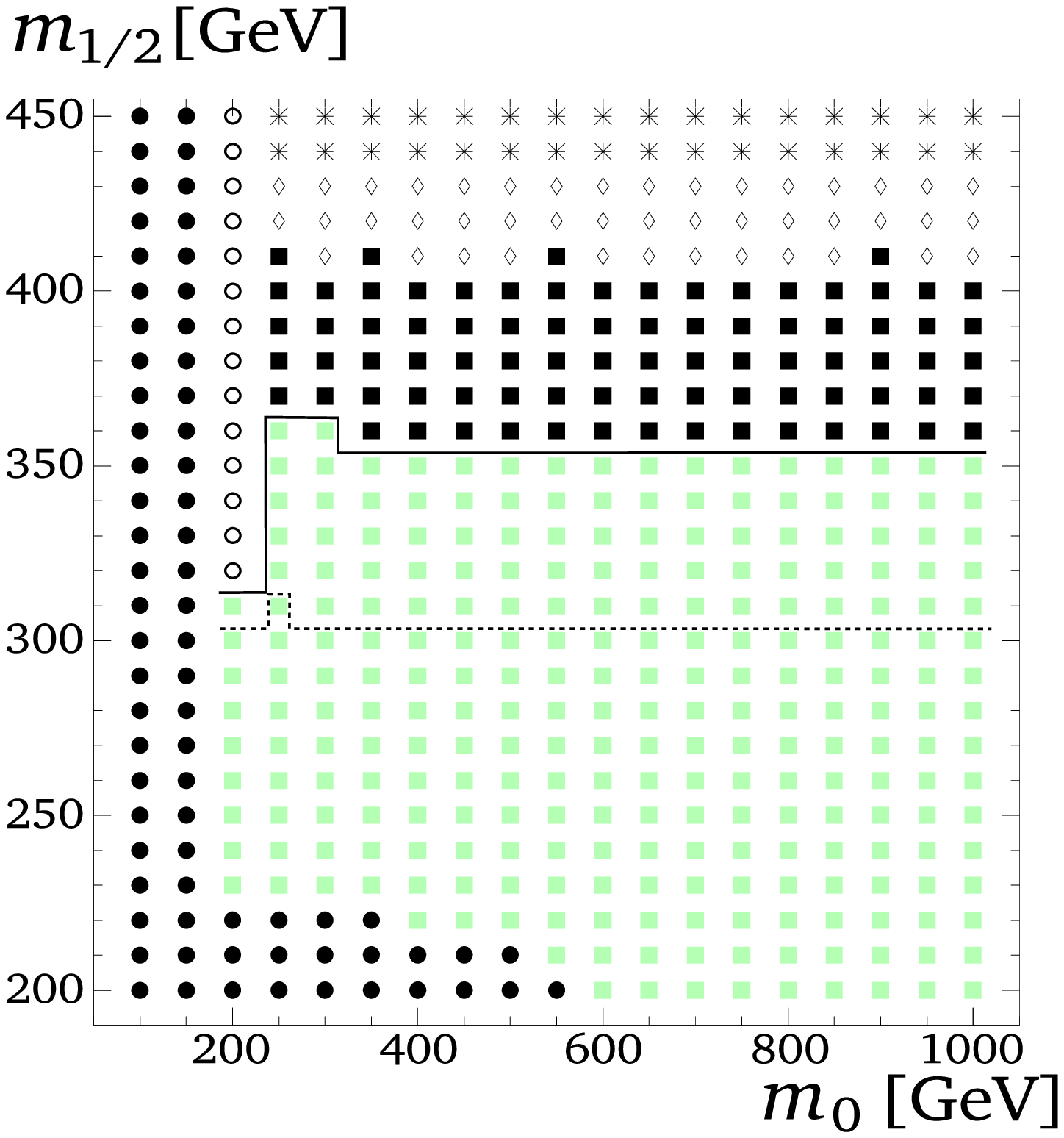}
 \rput(4,16){a)}
 \rput(12,16){b)}
 \rput(4,8){c)}
 \rput(12,8){d)}
 \caption{Reach of Fermilab Tevatron Run II using the displaced vertex signal
   in the $m_{1/2}\otimes m_0$ plane for $\mu > 0$.  In the points below the
   solid line the expected number of events ($NEv$) is greater than 4, while
   the region below the dashed line exhibits a sizeable statistics ($NEv>15$).
   The black squares denote points with an expected number of events greater
   than 4 for an integrated luminosity of 8 $\text{fb}^{-1}$, while the grey
   (green) squares indicates the points presenting a high statistics
   ($NEv>15$) for this luminosity.  Moreover, points denoted by diamonds have
   $2 < NEv<4$ with 8 $\text{fb}^{-1}$, while the stars correspond to points
   with $NEv<2$ for ${\cal L} =8\,\mathrm{fb}^{-1}$. The present direct search
   limits are violated at the points indicated with a round black circles,
   while the white circles mark the points where the neutralino is no longer
   the LSP.  The choice of the remaining parameters is the following: In (a)
   $\tan\beta=10$, $A_0 = -100\,$GeV, and the R--parity violating parameters
   for each mSUGRA point are such that they are compatible with neutrino data
   and the neutralino decay length is maximized (best case scenario). In (b)
   we kept the mSUGRA parameters as in (a), however we took the R--parity
   violating parameters for each mSUGRA point such that they are
   compatible with neutrino data and the neutralino decay length is minimized
   (worst case scenario).  In (c) the parameters are as in (a) but for
   $A_0=1000\,$GeV. Panel (d) was obtained using the parameters in (a) except
   for $\tan\beta=40$.}
\label{fig:reach}
\end{figure}

We are now set to study the discovery reach of the Tevatron using the
displaced vertex signal. In Fig.~\ref{fig:reach} we show the region of the
$m_0 \otimes m_{1/2}$ plane where the displaced vertex signal can be
established at Tevatron for integrated luminosities ${\cal L}$ of 2 fb$^{-1}$
and 8 fb$^{-1}$ and fixed values of $A_0$, $\tan\beta$, and
${\mathrm{sign}}(\mu)~(>0)$.  Our conventions are as follows: in the points
below the solid line the expected number of events ($NEv$) is greater than 4,
while the region below the dashed line exhibits a sizeable statistics
($NEv>15$).  The black squares denote points with an expected number of events
greater than 4 for an integrated luminosity of 8 $\text{fb}^{-1}$, while the
grey (green) squares indicates the points presenting a high statistics
($NEv>15$) for this luminosity.  Moreover, points denoted by diamonds have $2
< NEv<4$ with 8 $\text{fb}^{-1}$, while the stars correspond to points with
$NEv<2$ for ${\cal L} =8\,\mathrm{fb}^{-1}$. The present direct search limits
are violated at the points indicated with a round black circle, while the
white circles mark the points where the neutralino is no longer the LSP but
the lighter stau. The stau is to short-lived to produce a visible decay length
as has been shown in \cite{Hirsch:2002ys}. Therefore a different analysis is
required for such cases.

In Fig.~\ref{fig:reach}(a), top-left panel, we present the Tevatron reach for
$A_0= - 100$ GeV, $\tan\beta=10$ and $\mu > 0$, taking the the R--parity
violating parameters such that we obtain the largest decay length for the
lightest neutralino and being at the same time compatible at 3$\sigma$ level
with the present neutrino data.  We see from this figure that the Tevatron
with 2 fb$^{-1}$ integrated luminosity is able to probe $m_{1/2} \lesssim 350
$ GeV at moderate and large values of $m_0$.  At small values of $m_0$, where
the decay lengths are smaller and the main decay channel is $b \bar{b} \nu$,
the Tevatron can probe well the region $m_{1/2} \lesssim 310$ GeV. This reach
is larger than the one provided by trilepton and multilepton searches
\cite{Magro:2003zb}.  Moreover, for $m_{1/2} \lesssim 300$ GeV and $m_0
\gtrsim 300$ GeV we expect a large number of events, which can not only make
the discovery easy, but also be used to perform preliminary studies of the
neutralino branching ratios and confront them with the predictions of our
model. The Tevatron reach can be considerably extended provided larger
integrated luminosities become available, for instance, the region $m_{1/2}
\lesssim 410$ GeV and large values of $m_0$ can be explored for ${\cal L} = 8$
fb$^{-1}$ with the region $m_{1/2} \lesssim 350$ GeV exhibiting a large
statistics.

In order to understand the effect of different choices of the R--parity
violating parameters, we present in Fig.~\ref{fig:reach}(b), top-right panel,
the Tevatron reach for R--parity violating parameters leading to a minimal
decay length of the lightest neutralino compatible with neutrino data while
keeping the R-parity conserving parameters fixed. As expected from the
previous discussion, the reach is in both cases rather similar.

To study how our results depend upon $A_0$, we present in
Fig.~\ref{fig:reach}(c), lower-left panel, the displaced vertex discovery
reach at the Tevatron for $A_0 = - 900$ GeV and all other parameters as in
Fig.~\ref{fig:reach}(a) and the R--parity violating parameters such that the
neutralino decay length is maximized.  The changes in the predicted number of
events are again at the level of a few percent which leads to slightly smaller
Tevatron reach. Furthermore, there is a region with a stau LSP which cannot be
covered by the present analysis.  In Fig.~\ref{fig:reach}(d) the mSUGRA
parameters are chosen as in (a) but with $\tan\beta=40$.  Again we have chosen
the R--parity violating parameters such that the neutralino decay length is
maximized.  As we can see, the Tevatron reach increases at moderate $m_0$,
remaining approximately the same at large $m_0$. As in the previous case,
Fig.~\ref{fig:reach}(c), the stau is the LSP at small $m_0$ and thus this
region is not covered by our analysis.

We can learn from the Fig.~\ref{fig:reach} panels that the displaced
vertex signal remains approximately stable, showing that this channel
depends mainly upon $m_{1/2}$ and hardly on the other parameters,
including the R--parity violating ones provided they are in the range
allowed by neutrino data.  This is at variance with the behavior of
the trilepton and multilepton signals for our R--parity violating
SUGRA model~\cite{Magro:2003zb,Barger:1994nm}.  The reason for this
difference is that the decay length depends, in addition to
$m_{\tilde\chi^0_1}$, mainly upon the two ratios
$|\boldsymbol{\epsilon} / \mu|$ and $|\boldsymbol{\Lambda}|^2
/\mathrm{det}(m_{\tilde \chi^0})|$, which are constrained to lie in
narrow ranges by the neutrino data~\cite{Hirsch:2000ef}.  In contrast,
as seen in Fig.~\ref{fig:br}, the neutralino branching ratios into
different allowed channels exhibits a much stronger dependence on the
RmSUGRA parameters \cite{Magro:2003zb,Barger:1994nm} being, therefore,
less robust.

\begin{figure}[htbp]
  \includegraphics[scale=0.5]{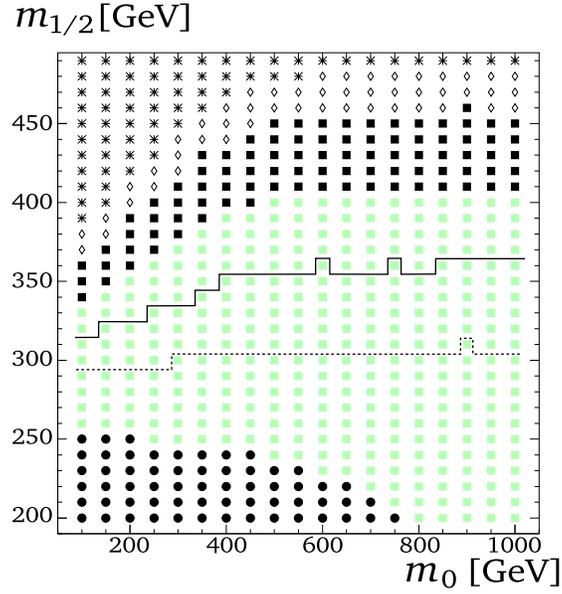} \caption{Reach of
    the Fermilab Tevatron Run II using the displaced vertex signal in
    the $m_{1/2}\otimes m_0$ plane. All the parameters are as in
    \protect{Fig.~\ref{fig:reach}}(a) except for an increase in the
    integrated luminosity from 8~fb${}^{-1}$ to $25\,
    \mathrm{fb}^{-1}$.}
  \label{fig:25fb-1}
\end{figure}

For illustrative purpose only, as the maximum luminosity expected for
the Tevatron RUN-II will be at most 8$\,\mathrm{fb}^{-1 }$, we present
in Fig.~\ref{fig:25fb-1} the expected reach in the displaced vertex
channel assuming an integrated luminosity of 25 fb$^{-1}$ in order to
allow easier comparison with the Tevatron reach of the trilepton
signal in the MSSM~\cite{tata1} and the trilepton and multilepton
signals with RmSUGRA obtained in Ref.~\cite{Magro:2003zb}. Comparing
the last two signals we learn that the presence of R--parity violation
enhances the Tevatron reach over most of the parameter space,
specially at moderate and large $m_0$.  Moreover, the reach in
$m_{1/2}$ of the displaced vertex signal is more than twice the one of
the trilepton and multilepton signals for basically all values of
$m_0$ and $\tan\beta$.

\begin{figure}[htbp]
  \includegraphics[scale=0.5]{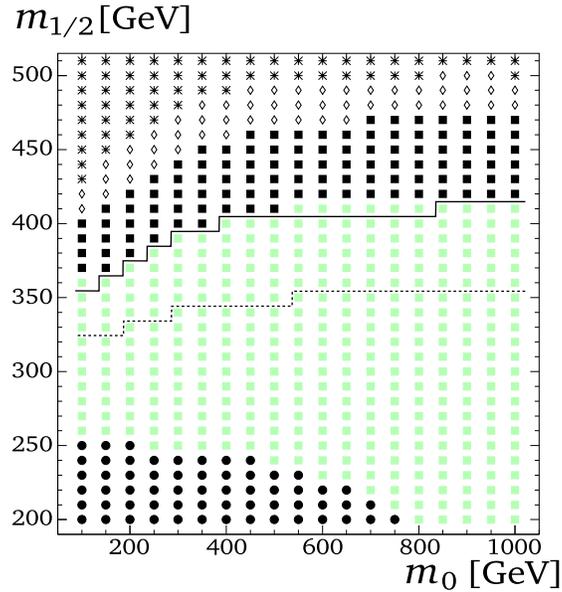} 
  \caption{Reach of Fermilab Tevatron Run II requiring the
    reconstruction of at least one displaced instead of two displaced
    vertices.  All the parameters are as in \protect{Fig.~\ref{fig:reach}}(a).}
  \label{fig:single}
\end{figure}

Despite the large reach of the two-displaced-vertices channel, it is
interesting to study whether we can further enlarge the Tevatron sensitivity
to RmSUGRA.  One possible way to achieve this goal is to consider events
containing one or more displaced vertices \cite{displacedvertices} that pass
our selection cuts instead of requiring two displaced vertices. We depicted in
Fig.~\ref{fig:single} the Tevatron reach requiring at least one reconstructed
vertex in the events for the same choice of parameters as in
Fig.~\ref{fig:reach}(a). As we can see, the reach in $m_{1/2}$ can be extended
in 50 GeV in accordance with ref.~\cite{displacedvertices}.  However, this
results should be taken with a pinch of salt since it is not obvious that our
cuts eliminate completely the SM backgrounds when there is just one observed
displaced vertex. To verify this result a detailed detector simulation is
called for.

%%%%%%%%%%%%%%%%%%%%%%%%%%%%%%%%%%%%%%%%%%%%%%%%%%%%%%%%%%%%%%%%%%%%%%
\section{Conclusion}
\label{concl}

We have analyzed displaced vertices from neutralino decays in a
 supergravity model with violation of R--parity (RmSUGRA) at
the Fermilab Tevatron. In this model R--parity is violated effectively
by bilinear terms in the superpotential and soft supersymmetry
breaking sector.  Despite the small R--parity violating couplings
needed to generate the neutrino masses and mixings indicated by
current neutrino data, the lightest supersymmetric particle 
can decay inside the detector.  This leads to a
phenomenology quite distinct from that of the R--parity conserving
scenario.  We have quantified the Tevatron reach for the displaced
vertices signal, displaying our results in the $m_0 \otimes m_{1/2}$
plane.

In contrast with other studies \cite{displacedvertices}, we have a background
free signal with a very stable yield which depends mainly on the neutralino
mass. This follows from our requirement of two displaced vertices instead of
only one \cite{displacedvertices}, and the fact that we imposed all
constraints of the neutrino data on the RmSUGRA parameters.  In fact, even for
$m_{1/2} \approx 470$~GeV ($m_{\tilde\chi_1^0}\approx215$~GeV), and moderate
to high $m_0$ values, a few events are expected for an integrated luminosity
of 8 fb${}^{-1}$, with a minor dependence in other RmSUGRA parameters.  This
reach is considerably larger than the one in the trilepton/multilepton
channels in either the MSSM or RmSUGRA.  Therefore, the displaced-vertices
signal is an essential tool to extend considerable the search possibilities of
Tevatron in the case of broken R--parity.

%%%%%%%%%%%%%%%%%%%%%%%%%%%%%%%%%%%%%%%%%%%%%%%%%%%%%%%%%%%%%%%%%%%%%%

\begin{acknowledgments}
  We thank A. Garcia-Bellido for discussions.  This work was supported by
  Spanish grant BFM2002-00345, by the European Commission Human Potential
  Program RTN network MRTN-CT-2004-503369, by Conselho Nacional de
  Desenvolvimento Cient\'{\i}fico e Tecnol\'ogico (CNPq), and by
  Funda\c{c}\~ao de Amparo \`a Pesquisa do Estado de S\~ao Paulo (FAPESP).
  W.~P. has been supported by a MCyT Ramon y Cajal contract and partly by the
  'Swiss Nationalfonds'.  D.R is supported by "Secretar\'{\i}a de Estado de
  Educaci\'on y Universidades'' postdoc contract SB2001-0056
\end{acknowledgments}

%%%%%%%%%%%%%%%%%%%%%%%%%%%%%%%%%%%%%%%%%%%%%%%%%%%%%%%%%%%%%%%%%%%%%%

%%%%%%%%%%%%%%%%%%%%%%%%%%%%%%%%%%%%%%%%%%%%%%%%%%%%%%%%%%%%%%%%%%%%%%
\newpage
%%%%%%%%%%%%%%%%%%%%%%%%%%%%%%%%%%%%%%%%%%%%%%%%%%%%%%%%%%%%%%%%%%%%%%

%%%%%%%%%%%%%%%%%%%%%%%%%%%%%%%%%%%%%%%%%%%%%

%%%%%%%%%%%%%%%%%%%%%%%%%%%%%%%%%%%%%%%%%%%%%

%%%%%%%%%%%%%%%%%%%%%%%%%%%%%%%%%%%%%%%%%%%%%

%%%%%%%%%%%%%%%%%%%%%%%%%%%%%%%%%%%%%%%%%%%%%

% \begin{figure}[htbp]
%   \includegraphics[scale=0.5]{nevdlm0400.eps}
%   \caption{Same as Fig.~\protect{\ref{fig:nevdl}} but fixing $m_0 =
%       400\,$GeV.  }
%   \label{fig:nevdl0400}
% \end{figure}
% 

%%%%%%%%%%%%%%%%%%%%%%%%%%%%%%%%%%%%%%%%%%%%%
%%%%%%%%%%%%%%%%%%%%%%%%%%%%%%%%%%%%%%%%%%%%%

%%%%%%%%%%%%%%%%%%%%%%%%%%%%%%%%%%%%%%%%%%%%%

%%%%%%%%%%%%%%%%%%%%%%%%%%%%%%%%%%%%%%%%%%%%%

%%%%%%%%%%%%%%%%%%%%%%%%%%%%%%%%%%%%%%%%%%%%%%%%%%%%%%%%%%%%%%%%%%%%%%
\end{document}